\documentclass[aps,twocolumn]{revtex4}

\usepackage{amsmath}
\usepackage{graphicx}
\usepackage[colorlinks=true, allcolors=blue]{hyperref}

\begin{document}
\title{Simulations of stochastic fluid dynamics near a critical point 
in the phase diagram}

\author{Chandrodoy Chattopadhyay, 
Josh Ott, 
Thomas Sch\"afer,  
and Vladimir V. Skokov}
\affiliation{Department of Physics, 
North Carolina State University,
Raleigh, NC 27695}

\begin{abstract}
 We present simulations of stochastic fluid dynamics
in the vicinity of a critical endpoint belonging to the
universality class of the Ising model. This study is 
motivated by the challenge of modeling the dynamics of
critical fluctuations near a conjectured critical endpoint
in the phase diagram of Quantum Chromodynamics (QCD). We
focus on the interaction of shear modes with a conserved
scalar density, which is known as model H. We show that 
the observed dynamical scaling behavior depends on the 
correlation length and the shear viscosity of the fluid.
As the correlation length is increased or the viscosity 
is decreased we observe a cross-over from the dynamical
exponent of critical diffusion, $z\simeq 4$, to the 
expected scaling exponent of model H, $z\simeq 3$. We 
use our method to investigate time-dependent correlation
function of non-Gaussian moments $M^n(t)$ of the order 
parameter. We find that the relaxation time depends in 
non-trivial manner on the power $n$.
\end{abstract}
\maketitle

\section{Introduction}

 The transition from hadronic matter to a quark gluon plasma
along the finite temperature axis in the QCD phase diagram is 
known to be a smooth crossover \cite{Aoki:2006we}. As the baryon 
doping is increased, this crossover may turn into a first-order 
phase transition at a critical endpoint \cite{Stephanov:1998dy}. 
Experimental searches for critical behavior have focused on a 
possible non-monotonic dependence of fluctuation observables, 
such as the cumulants of conserved charges, on the beam energy
\cite{Bzdak:2019pkr,Bluhm:2020mpc,An:2021wof,STAR:2020tga}. 
Understanding the dynamical evolution of these observables in 
a heavy ion collision requires a hydrodynamic theory that
incorporates the effects of fluctuations \cite{Rajagopal:1992qz,
Berdnikov:1999ph,Son:2004iv}.

 Theories of this type were first classified by Hohenberg and
Halperin \cite{Hohenberg:1977ym}. They include purely relaxational
dynamics (model A) \cite{Schweitzer:2020noq,Schaefer:2022bfm}, 
the diffusive dynamics of a conserved charge (model B)
\cite{Berdnikov:1999ph,Nahrgang:2018afz,Schweitzer:2021iqk}, 
and the diffusive evolution of an order parameter field advected
by the momentum density of the fluid (model H). Model H is
expected to govern the dynamics near a possible critical endpoint
in the QCD phase diagram \cite{Son:2004iv}.

Stochastic hydrodynamic theories have been studied using a 
variety of methods \cite{Mukherjee:2015swa,Akamatsu:2016llw,
Stephanov:2017ghc,Martinez:2018wia,Akamatsu:2018vjr,An:2019osr,
An:2019csj,An:2020vri}, but there is little work on direct
numerical simulation (see \cite{Berges:2009jz,Nahrgang:2018afz,
Schweitzer:2020noq,Schweitzer:2021iqk,Pihan:2022xcl,
Schaefer:2022bfm} for exceptions). In particular, model H has 
not been studied numerically. This is related to the fact 
that numerical simulations face a number of obstacles, including 
the need to regularize and renormalize short-distance noise, the
requirement to implement fluctuation-dissipation relations, and
the necessity to resolve ambiguities in the definition of
stochastic partial differential equations. 

 In the present work, we describe a numerical implementation of 
model H using a Metropolis method previously applied to models
A, B and G (chiral dynamics) \cite{Florio:2021jlx,Schaefer:2022bfm,
Chattopadhyay:2023jfm,Florio:2023kmy}. The novel feature of model H 
compared to purely relaxational or diffusive theories is the presence 
of ``mode couplings" or ``Poisson-brackets". These terms describe 
advective interactions that conserve the hydrodynamic Hamiltonian,
but lead to non-linear mode couplings between shear waves and the
diffusive evolution of the order parameter. In the following, we
introduce the model, explain our numerical approach,  present a
number of consistency checks, and then present results for the 
dynamical evolution of non-Gaussian moments. We comment on other
applications and possible extensions of our methods. 

\section{Model H}

 Model H is defined by \cite{Hohenberg:1977ym,Folk:2006ve} 
\begin{align}
\label{modH_1}
\partial_t\phi &=   \Gamma\,\nabla^2 
        \left(\frac{\delta{\cal H}}{\delta \phi}\right)
- \left(\vec\nabla\phi\right) \frac{\delta{\cal H}}{\delta \vec{\pi}_T}
      + \zeta , \\
\label{modH_2}
\partial_t \pi^T_i &= \eta \nabla^2 
    \left(\frac{\delta{\cal H}}{\delta \pi^T_i}\right)
    +  P^T_{ij} 
    \left[\left(\nabla_j\phi\right) \frac{\delta{\cal H}}{\delta\phi} \right] \nonumber \\
    & \hspace{1.5cm}
    -  P^T_{ij} \left[ 
        \nabla_k\left( \pi^{T}_j \frac{\delta{\cal H}}{\delta
       \pi^{T}_k}\right)
     \right] + \xi_i\, , 
\end{align}
where $\phi$ is the order parameter density, $\vec{\pi}$ is the
momentum density of the fluid, $\Gamma$ and $\eta$ are transport
coefficients \footnote{
We note that some authors, including Hohenberg and Halperin, 
define model H without the self-coupling of $\vec{\pi}^T$, based
on the observation that this coupling is irrelevant in the 
sense of the renormalization group. We will refer to this 
truncation as model H0.}. 
We can take $\phi$ to be proportional to the
specific entropy $s/n$ of the fluid \cite{Akamatsu:2018vjr,
An:2022jgc}. $\Gamma$ is the thermal diffusivity, and $\eta$ 
is the shear viscosity. The transverse projection operator is
given by 
\begin{align}
    P^T_{ij} = \delta_{ij}- \frac{\nabla_i\nabla_j}{\nabla^2} 
\end{align}
and $\pi^T_i=P^T_{ij}\pi_j$. The Hamiltonian (the free energy 
functional) is given by 
\begin{equation}
\label{H_Ising}
    {\cal H}  = \int d^dx \left[ \frac{1}{2\rho} (\vec{\pi}_T)^2
    +  \frac{1}{2} (\nabla \phi)^2  +  \frac{m_0^2}{2} \phi^2 
    +   \frac{\lambda}{4}  \phi^4  - h \phi\right]  ,
\end{equation}
where $\rho$ is the mass density (the density of enthalpy in the 
relativistic case), $m_0$ is the bare inverse correlation length, 
$\lambda$ is a non-linear self-coupling, and $h$ is an external
field. In practical applications, these parameter can be mapped 
onto the chemical potential-temperature $(\mu,T)$ plane, see for 
example \cite{Parotto:2018pwx,Kahangirwe:2024cny}. The noise 
terms $\zeta$ and $\xi_i$ are random fields constrained by
fluctuation-dissipation relations. The noise correlation functions 
are given by  
\begin{align}
    \langle \zeta (t, \vec{x}) \zeta (t', \vec{x}') \rangle &= 
    -2 T\, \Gamma\, \nabla^2 \delta(\vec{x}-\vec{x}')\delta(t-t')\, ,\\
  \langle \xi_i (t, \vec{x}) \xi_j (t', \vec{x}') \rangle &= 
    -2 T\, \eta\, P^T_{ij} \nabla^2
      \delta(\vec{x}-\vec{x}')\delta(t-t')\, \label{noise_correlator2} .  
\end{align}
Note that equations (\ref{modH_1},\ref{modH_2}) describe the 
interaction of shear modes with the order parameter, but they do
not include sound modes. This truncation is expected to be 
sufficient to describe the critical dynamics of the fluid
\cite{Hohenberg:1977ym,Son:2004iv}, but for other applications 
it will be interesting to include the coupling to  longitudinal modes \cite{Martinez:2019bsn}.

\section{Numerical method}

In order to study the theory numerically, we discretize the fields 
$\phi(\vec{x})$ and $\vec\pi(\vec{x})$ on a $d$-dimensional 
lattice $\vec{x}=\vec{n}a$ with $n_i=1,\ldots,N$. In the 
following, we will focus on $d=3$. The main idea underlying the
algorithm we employ is that the dissipative and stochastic updates 
are combined into a single Metropolis step. This method ensures 
that fluctuation-dissipation relations are satisfied and 
that the fluid equilibrates to a state in which the fields 
$\phi(\vec{x})$ and $\vec\pi^T(\vec{x})$ are sampled from the
distribution $\exp(-{\cal H}/T)$. The Metropolis step is 
followed by a deterministic step that implements the 
non-dissipative mode coupling terms. 

 The Metropolis update for the field $\phi$ is given by 
\begin{align}
\phi^{\rm trial}(\vec{x},t+\Delta t) & = \phi(\vec{x},t) 
   + q^{(\mu)}\, ,\\
\phi^{\rm trial}(\vec{x}+\hat\mu,t+\Delta t) &= 
   \phi(\vec{x}+\hat{\mu},t)  - q^{(\mu)}\,  , \\
 q^{(\mu)} &= \sqrt{2\Gamma T(\Delta t)}\, \zeta^{(\mu)}  \, ,
\label{phi-stoch}
\end{align}
where each $\zeta^{(\mu)}$ is a Gaussian random variable with zero mean 
and unit variance, and $\hat{\mu}$ is an elementary lattice vector 
in the direction $\mu=1,\ldots,d$. The update is accepted with
probability ${\rm min}(1,e^{-\Delta{\cal H}/T})$. This algorithm 
is based on the observation that the average update $\langle
[\phi(\vec{x},t+\Delta t) - \phi(\vec{x},t)]\rangle$ realizes 
the diffusion equation, and the second moment $\langle [\phi
(\vec{x},t+\Delta t)-\phi(\vec{x},t)]^2\rangle$ reproduces the 
noise term, see \cite{Florio:2021jlx,Chattopadhyay:2023jfm}.

We can follow the same procedure for $\vec\pi$ and perform a
trial update 
\begin{align}
\pi_\nu^{\rm trial}(\vec{x},t+\Delta t) & = 
         \pi^T_\nu(\vec{x},t) + r_{\nu}^{(\mu)}\, ,\\
\pi_\nu^{\rm trial}(\vec{x}+\hat\mu,t+\Delta t) &= 
  \pi^T_\nu(\vec{x}+\hat{\mu},t)  - r_{\nu}^{(\mu)}\,  ,\\
 r_{\nu}^{(\mu)} &= \sqrt{2\eta T (\Delta t)}\, \xi_{\nu}^{(\mu)}  \, , 
\label{pi-stoch}
\end{align}
where $r_{\nu}^{(\mu)}$ is a random flux and $\xi_\nu^{(\mu)}$ are 
Gaussian random variables with $\langle \xi_\mu^{(\alpha)}\xi_\nu^{(\beta)}
\rangle=\delta_{\mu\nu}\delta^{\alpha\beta}$. Again, the update is
accepted with probability ${\rm min}(1,e^{-\Delta{\cal H}/T})$. 
After a sweep through the lattice we project on the 
transverse component of the momentum density, $\pi_\mu^T (\vec{x},t)
=P^T_{\mu\nu} \pi_\nu(\vec{x})$. The projection is carried out 
in Fourier space. 

 The deterministic update implements the advection terms 
\begin{align}
\label{phi_adv}
\partial_t\phi &= -\frac{1}{\rho}\pi^T_\mu\nabla_\mu \phi \, ,\\
\label{pi_adv}
\partial_t\pi^T_\mu &= -\frac{1}{\rho}\pi^T_\nu\nabla_\nu \pi^T_\mu
  - P_{\mu\nu}^T\left[ (\nabla_\nu \phi) \nabla^2\phi \right]\, .
\end{align}
In the continuum limit, these equations conserve the integrals 
of $\phi$ and $\pi_\mu^T$, as well as the Hamiltonian ${\cal H}$.
We have found that it is important to preserve these conservation
laws in the lattice theory to the greatest extent possible. 
Using the skew discretized derivatives introduced by Morinishi
et al.~\cite{Morinishi:1998} it is possible to construct an 
advection step that conserves integrals of the kinetic energy 
terms $\frac{1}{2\rho}(\pi^T_\mu)^2+\frac{1}{2} (\nabla_\mu\phi)^2$ 
exactly \cite{Chattopadhyay:2024}. We integrate the equations
(\ref{phi_adv},\ref{pi_adv}) using the strongly stable third-order
Runge-Kutta scheme of Shu and Osher \cite{Shu:1988}. The fields 
$\phi$ and $\pi_\mu^T$ satisfy conservation laws after projection,
and the total energy is conserved to very good accuracy. This 
statement can be quantified in terms of the observed shift in the
critical mass (see Sect.~\ref{sec:res}), which is about 1\%. 

  A complete update consists of a Metropolis update of all 
fields, followed by a Runge-Kutta step for the advection terms. 
After every update of the momentum density, the projector $P^T$ 
is applied in Fourier space. The time step $\Delta t$ is chosen 
such that the acceptance rate of the Metropolis step is of order 
$1/2$. In practice, we have used $\Delta t=0.04/\Gamma$.

\section{Results}
\label{sec:res}

  We have solved the model H equations on a periodic lattice 
of size $L^3$ with $L=Na$. In the following, we will set $a=1$, 
which means that all quantities that have units of length are 
measured in units of $a$. We will also set $\Gamma=1$, which 
implies that our unit of time is $a^4/\Gamma$. We tune $m^2$ 
to fix the correlation length in units of $a$. In particular, 
for $h=0$ there is a critical $m^2 \equiv m_c^2$ at which the
correlation length diverges. In the following, we will study the 
dependence on the parameters $\eta$ and $\rho$. 

\begin{figure}[t]
\begin{center}
\includegraphics[width=0.9\columnwidth]{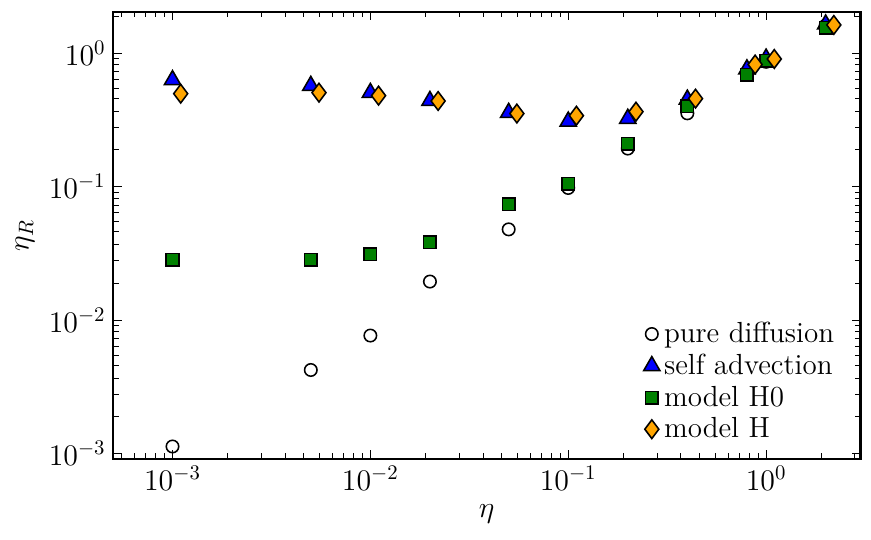}
\end{center}
\caption{Renormalized viscosity as a function of the bare 
viscosity in four different models: 1) pure momentum diffusion 
(no mode-couplings), 2) self-advection ($\pi^T$ only couples to
itself), 3) critical model H0 (mutual advection of $\phi$ and 
$\vec\pi_T$ only), and 4) model H at the critical point. All
data were taken in lattice volume $V=L^3$ with $L=48$ and 
with $\rho=1$. Model H results were offset horizontally for better visibility.
\label{fig:eta-eff}}
\end{figure}

 {\it Static behavior:} We first consider the static equilibrium 
state of the fluid described by Eqs.~(\ref{modH_1},\ref{modH_2}).
The Hamiltonian does not contain any coupling between $\phi$ and 
$\pi_T$, and in the continuum, we expect the static correlation 
function $C(\vec{x})=\langle\phi(0,t)\phi(\vec{x},t)\rangle$ of 
the order parameter to be unaffected by the dynamics of the momentum  
density. In particular, the critical value of the mass parameter 
is expected to agree with the value previously determined in model 
A and B, $m_c^2=-2.2858$ (for $\lambda=4$)\cite{Schaefer:2022bfm,
Chattopadhyay:2023jfm} and the correlation function $C(\vec{x})$ 
is predicted to be the same in model B and model H \footnote{
The correlation function in model A differs from that in model 
B and H because of the effects of conservation laws in a finite 
volume.}.
In practice, our advection step does not conserve the potential 
energy of $\phi$ exactly, and we see a small shift $\delta m_c^2
=-0.030$ in the critical mass parameter \footnote{
A similar shift was observed in the model G calculation described 
in \cite{Florio:2021jlx}.}.
After taking this shift into account, we find that the correlation
function $C(\vec{x})$ is the same in models B and H 
\cite{Chattopadhyay:2024}.

{\it Renormalization of the shear viscosity:} Next we study the 
dynamics of the momentum density without the coupling to the 
order parameter $\phi$. This corresponds to the non-critical 
dynamics of a fluid in the limit $m^2\to\infty$. It was previously
observed that the non-linear self-coupling of $\vec{\pi}$ leads a 
renormalization of the shear viscosity, referred to as the
``stickiness of sound'' in \cite{Kovtun:2011np}. In the present 
case, the phenomenon is more accurately characterized as the
stickiness of shear waves. A one-loop calculation predicts that 
\cite{Kovtun:2011np,Chafin:2012eq}
\begin{align}
\label{eta-R}
\eta_R = \eta + \frac{7}{60\pi^2} \frac{\rho T\Lambda}{\eta}\, ,
\end{align}
where $\Lambda\simeq \pi/a$ is the UV cutoff. We have extracted the 
renormalized viscosity from the exponential decay of the unequal 
time correlation function $\langle \pi^T_i(0,\vec{k})\pi^T_i(t,
-\vec{k})\rangle \sim \exp(-(\eta_R/\rho) k^2 t)$ for the first 
non-tivial momentum mode in the cartesian direction $j$ for 
$i\neq j$. The result is shown in Fig.~\ref{fig:eta-eff}. We 
observe that as the bare viscosity is reduced, the renormalized 
$\eta_R$ levels off and then increases, in agreement with 
Eq.~(\ref{eta-R}).

\begin{figure}[t]
\begin{center}
\includegraphics[width=0.95\columnwidth]{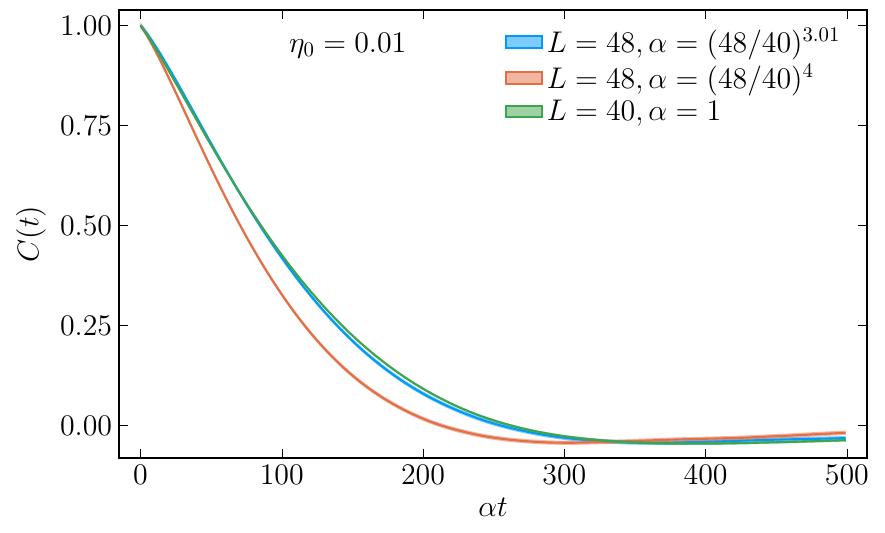}
\end{center}
\caption{
Dynamic order parameter correlation function $C(t,\vec{k})$ for the 
second non-trivial momentum mode and for different values of $L$ 
plotted as a function of the scaled time variable. This figure shows 
data taken at $\eta=10^{-2}$ for $L=40$ and $48$. Data collapse 
occurs for $z\simeq 3.01$, and the model B value $z=4$ is clearly 
excluded.
\label{fig:dyn-scal}}
\end{figure}

 We have also studied the renormalization of $\eta$ in a theory 
in which the coupling between $\phi$ and $\vec\pi^T$ is retained, 
but the self-advection of $\vec\pi^T$ is ignored. This is a 
consistent truncation of model H, which we will call model H0. 
Indeed, the self-coupling of $\vec\pi^T$ is irrelevant in the 
sense of the renormalization group (RG), and model H0 is sufficient 
to compute critical exponents for the liquid-gas critical endpoint
\cite{Hohenberg:1977ym}. In model H0 critical fluctuations lead to 
a multiplicative renormalization 
\begin{align}
\label{eta-crit}
\eta_R = \eta \left[ 1 + \frac{8}{15\pi^2}\log(\xi/\xi_0)\right], 
\end{align}
where $\xi$ is the correlation length and $\xi_0\simeq a$ is the 
bare correlation length. This effect is difficult to observe 
because of the small prefactor in Eq.~(\ref{eta-crit}). Non-critical
fluctuations generate a finite additive renormalization, $\eta_R
=(T\xi_0)/(160\pi\Gamma)$. This effect is also much smaller compared 
to Eq.~(\ref{eta-R}). Indeed, Fig.~\ref{fig:eta-eff} shows that 
the renormalized viscosity in model H0 continues to drop with 
$\eta$ and can reach very small values of order $\eta_R\simeq 
10^{-2}$.

{\it Dynamical scaling:} Consider the time-dependent correlation
function $C(\vec{k},t)=\langle \phi(\vec{k},0)\phi(-\vec{k},t)
\rangle$. Dynamical scaling is the hypothesis that $\tilde{C}
(k\xi,t/\xi^z)$, where $z$ is called the dynamical exponent, is 
a universal function near the critical point. To understand 
the behavior of the correlation function it is useful to start from 
the prediction of the mode coupling theory \cite{Hohenberg:1977ym}. 
In this approximation, it is assumed that the correlation function
is controlled by a single relaxation rate, $C(\vec{k},t)\sim\exp(-
\Gamma_k t)$, and that the renormalized viscosity $\eta_R$ is 
a constant, independent of $\xi$. Based on these assumptions
one finds 
\begin{align}
\label{Kawasaki}
   \Gamma_k = \frac{\Gamma}{\xi^4} 
      \left(k\xi\right)^2 \left(1+(k\xi)^2\right)
     + \frac{T}{6\pi\eta_R\xi^3} K(k\xi)\, , 
\end{align}
where the Kawasaki function $K(x)$ is given by $K(x) \simeq x^2$ for
$x\ll 1$ and $K(x)\simeq (3\pi/8)x^3$ for $x\gg 1$ \cite{Kawasaki:1970}.
This result suggests that the dynamic exponent crosses over from 
$z\simeq 4$ at modest values of the correlation length to $z\simeq 3$
if the correlation length is large, $\xi\gg (6\pi\eta_R\Gamma)/T$.

\begin{figure}[t]
\begin{center}
\includegraphics[width=0.95\columnwidth]{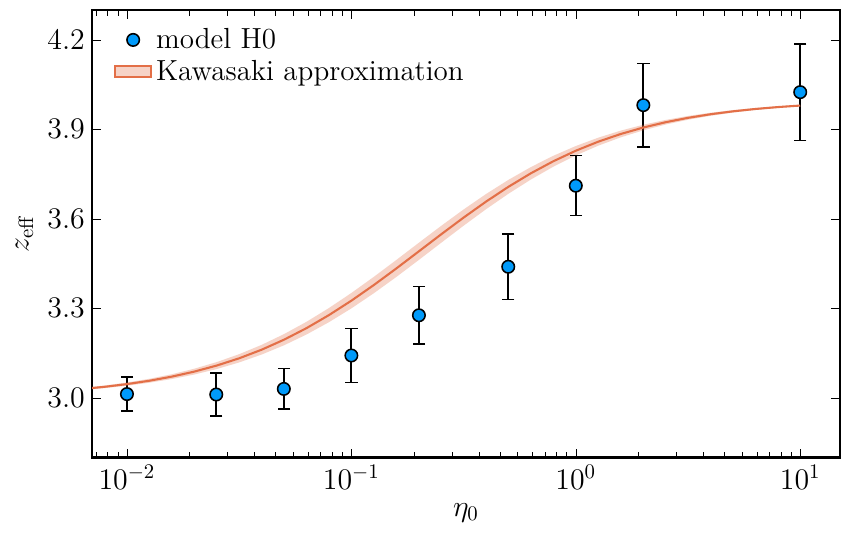}
\end{center}
\caption{Scaling exponent $z$ extracted from the dynamic correlation 
functions for different values of the 
bare viscosity $\eta$ in model H0. We determined $z$ by comparing
the correlation function for two different volumes, $L=40$ and $48$.  
We also show the prediction of the Kawasaki approximation,
Eq.~(\ref{Kawasaki}); the error band is defined by varying the correlation 
ength in the range $\xi \in [L/2\pi, L/2]$.  
\label{fig:z-eff}}
\end{figure}

The values $z=3$ and $4$ are approximate, a more sophisticated
calculation using the $\epsilon$-expansion gives $z=3.07$  and
$z=3.96$, but the Kawasaki function is a very good approximation
to the behavior of real fluids \cite{Swinney:1973}. We conclude from Eq.~(\ref{Kawasaki})
that in any finite volume we are likely to observe a scaling 
exponent between 3 and 4, and that observing the model H
scaling exponent requires a combination of a very large correlation
length and a small viscosity.

  Since the renormalized viscosity is smaller in model H0 (see
Fig.~\ref{fig:eta-eff}) we have explored the scaling behavior in 
model H0. At the critical point $m^2=m_c^2$ we compute the correlation
for a range of values of $\eta$ and $L^3$. For any given $\eta$
we look for data collapse when comparing different $L$ in order to 
determine the value of $z$. This is shown in Fig.~\ref{fig:dyn-scal} 
for $\eta=0.01$ and $L=40$ and $L=48$. We observe that scaling works 
very well. We then plot the extracted value of $z$ as a function of 
$\eta$ \footnote{The data in Fig.~\ref{fig:z-eff} are obtained by 
minimizing $|C(t, L\!=\!40) - C(\alpha t, L\!=\!48)|$ with respect 
to $\alpha=(40/48)^z$ in the regime $C(t)>0.15$ (with $C(0)\equiv 1$). 
The errorbar in $z$ is determined by propagating the errors in $C(t)$.}. 
The result is shown in Fig.~\ref{fig:z-eff}, and compared 
to the Kawasaki prediction. We observe the expected crossover from 
$z\simeq 4$ at large viscosity to $z\simeq 3$ for small viscosity. 
For $\eta=10^{-2}$ we obtain the dynamical exponent $z\simeq 3.013\pm0.058$,
consistent with  the prediction of the two-loop $\epsilon$-expansion, 
$z\simeq 3.0712$~\cite{adzhemyan1999h}.

 {\it Non-Gaussian moments:} Having established that our numerical 
results are compatible with theoretical expectations, we turn to 
an observable that is not easily predicted by approximate analytical 
methods. Consider the correlation function of higher moments of the 
order parameter
\begin{align}
  G_n(t) = \langle M^n(t)M^n(0)\rangle\, , \hspace{0.25cm}
  M(t) =\int_Vd^3x\, \phi(\vec{x},t)\, , 
\end{align}
where $V$ is a sub-volume of the simulation volume. Higher 
cumulants of the order parameter have been proposed as signatures
of critical behavior \cite{Stephanov:2008qz}, and their time 
evolution was previously studied in \cite{Mukherjee:2015swa,
An:2022jgc,Schaefer:2022bfm}.

\begin{figure}[t]
\begin{center}
\includegraphics[width=0.95\columnwidth]{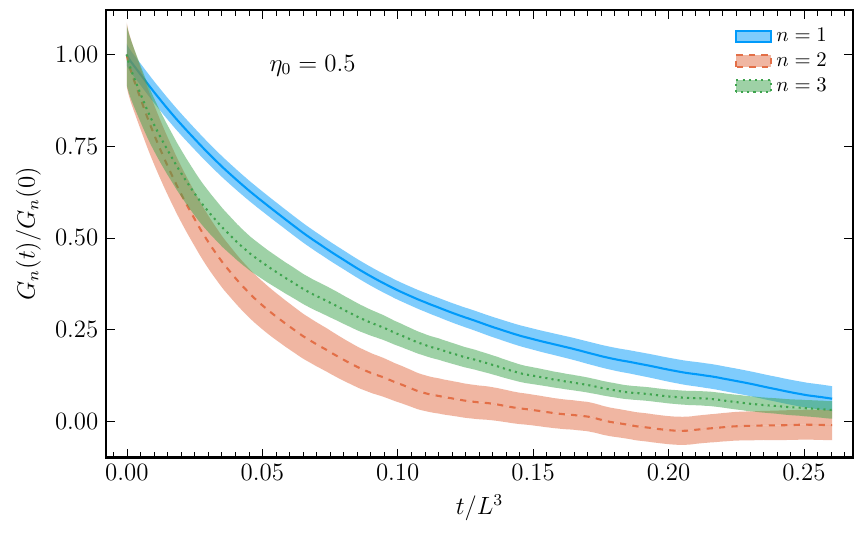}
\end{center}
\caption{
\label{fig:cumulants}
Correlation functions $G_n(t)=\langle M^n(t)M^n(0)\rangle$
in model H at the critical point for physical values of 
the parameters (see text). The calculation was performed for
$L=48$ with $M$ integrated over half of the volume.
For even $n$ we subtract disconnected pieces.}
\end{figure}

 We consider model H with parameters relevant to a QCD critical
endpoint. We take $a=0.75$ fm and $\Delta t=0.3$ fm, and consider
a temperature $T=130$ MeV. Then, the enthalpy density of a 
non-interacting quark-gluon plasma corresponds to $\rho=11.1$ 
in lattice units. A viscosity to entropy density ratio $\eta/s
=1/(4\pi)$ implies $\eta=0.50$ \footnote{
Note that $\eta$ is somewhat bigger than the minimum value of 
$\eta_R$ shown in Fig.~\ref{fig:eta-eff}. This is consistent 
with the observation by Kovtun et al.~\cite{Kovtun:2011np} that 
fluctuations lead to a bound on $\eta$ which is slightly weaker 
than the string theory bound $\eta/s\geq 1/(4\pi)$.}. 
We take a simulation volume $V_0=L^3$ and measure the order 
parameter in half the simulation volume. The results are shown in 
Fig.~\ref{fig:cumulants}. The correlation functions satisfy
dynamical scaling for all values of $n$, but the relaxation time 
depends on $n$. In particular, $G_2(t)$ decays more quickly than 
$G_1(t)$, but the relaxation rate of $G_3(t)$ is intermediate 
between $G_1(t)$ and $G_2(t)$. These results are not compatible 
with simple mean field models. Note that at the critical point 
the correlation length is only limited by $L$, and equilibration
is extremely slow: In physical units, the $1/e$ decay time in 
Fig.~\ref{fig:cumulants} exceeds $10^3$ fm.

\section{Summary and outlook:}

 In this work we have presented a method for performing stochastic
fluid dynamics simulations. We find that the dynamic scaling exponent
in a near-critical fluid depends sensitively on the value of the 
correlation length and the shear viscosity. Genuine Model H behavior 
with $z\simeq 3$ requires a large correlation length and small shear
viscosity. We also observe that while the self-coupling of the 
momentum density is technically irrelevant in the sense of the
renormalization group (the critical behavior of model H and H0 
is the same) it is numerically quite important in limiting how small
the viscosity can become. 

  We have studied the time evolution of higher moments of the 
order parameter, but we have not attempted to perform a complete
calculation that can be compared to higher order cumulants measured
in relativistic heavy ion collisions. For this purpose we have to 
couple our calculation to a realistic background that describes 
an expanding and cooling fluid. There are two basic approaches that 
one might follow in order to achieve this. One approach is to 
extend the methods described in this work to stochastic relativistic
fluid dynamics, where we retain all the degrees of freedom of the 
fluid (both shear and sound modes). This could be accomplished 
along the lines recently proposed in \cite{Basar:2024qxd}. Another 
option is to couple the model described here to a deterministic 
background flow, obtained from conventional fluid dynamic simulations. 
 
{\it Acknowledgments:}
This work is supported by the U.S. Department of Energy, Office 
of Nuclear Physics through the Contracts DE-FG02-03ER41260 (T.S.) 
and DE-SC0020081 (V.S.). We used computing resources provided by 
the NC State University High Performance Computing Services Core 
Facility (RRID:SCR-022168), as well as resources funded by the 
Wesley O.~Doggett endowment. We thank Andrew Petersen for assistance 
in working with the HPC infrastructure. 

\bibliographystyle{apsrev4-1}
\bibliography{bib}
\end{document}